# Self-organized synchronization of phonon lasers

Jiteng Sheng[1,2], Xinrui Wei[1], Cheng Yang[1], and Haibin Wu[1,2,*]

[1]State Key Laboratory of Precision Spectroscopy, East China Normal University, Shanghai 200062, China

[2]Collaborative Innovation Center of Extreme Optics, Shanxi University, Taiyuan 030006, China

*hbwu@phy.ecun.edu.cn

**Self-organized synchronization is a ubiquitous collective phenomenon, in which each unit adjusts their rhythms to achieve synchrony through mutual interactions. The optomechanical systems, due to their inherently engineerable nonlinearities, provide an ideal platform to study self-organized synchronization. Here, we demonstrate the self-organized synchronization of phonon lasers in a two-membrane-in-the-middle optomechanical system. The probe of individual membrane enables to monitor the real-time transient dynamics of synchronization, which reveal that the system enters into the synchronization regime via torus birth bifurcation line. The phase-locking phenomenon and the transition between in-phase and anti-phase regimes are directly observed. Moreover, such a system greatly facilitate the controllable synchronous states, and consequently a phononic memory is realized by tuning the system parameters. This result is an important step towards the future studies of many-body collective behaviors in multiresonator optomechanics with long distances, and might find potential applications in quantum information processing and complex networks.**

Synchronization plays an important role in many aspects of modern science. A well-recognized example in laser physics is the so-called injection locking [1,2]. It has also shown great potential in various laser-based technologies. Synchronization of optical clocks is essential in ultra-precise navigation, sensing, and time-keeping [3]. Synchronization of semiconductor laser arrays can lead to ultrahigh coherent power [4,5]. Synchronization of chaotic lasers is useful in chaos-based secure communications [6,7], and synchronization of superradiant lasers provides extremely high purity of frequency [8].

Phonon laser is the counterpart of conventional laser or photon laser. The lasing is defined as that the mechanical oscillator is driven into the self-oscillation regime when the driving power is above a threshold and the effective damping rate becomes negative. This field with single optomechanical oscillator has been extensively studied in the last few years [9-11]. As the development of nano- and microfabrication, multi-mechanical-oscillators recently become an active field. The realization of synchronous phonon lasers in such systems will not only find applications in quantum communications and complex networks, but also can explore many rich and fascinating nonlinear and quantum phenomena, such as chaotic and quantum synchronization [12-15]. A promising candidate to study synchronization of phonon lasers are multimode and multiresonator optomechanical systems, e.g., coupled microdisks [16], optomechanical crystals [17-19], hybrid microwave circuits [20,21], and multiple dielectric membranes in a Fabry-Perot cavity [22-29].

Recently, synchronization has been observed in the coupled micro-disks and nanomechanical oscillators with an optical racetrack cavity [30-33]. However, there is no such demonstration for the multi-membrane system. Self-organized synchronization of phonon lasers remains to be largely explored experimentally. Especially, dynamic control of phonon lasers via synchronization, which could be an essential ingredient for phonon-photon hybrid circuits, has remained elusive. In this work, we investigate the synchronization effect in an optomechanical system of two long-distance dielectric membranes interacting with one common intracavity field through dynamical backaction. We demonstrate the synchronous phonon lasers with the evidence of frequency entrainment, as well as the direct observation of phase locking. The real-time dynamics is studied by monitoring the motions of two membranes individually and simultaneously. Consequently, the route to synchronization is experimentally investigated in an optomechanical system for the first time. The transition between the regimes of in-phase and anti-phase synchronization is also observed. Moreover, synchronization is utilized as a useful resource to control the synchronous frequency by tuning the system parameters and this characteristic is exploited to realize a phononic memory, which could be straightforwardly extended to multichannel phononic switches and logic gates.

The schematic of the experimental setup is shown in Fig. 1. Two stoichiometric silicon nitride membranes separated by 6 cm are placed inside a Fabry-Perot cavity. The driving beam interacts with two membranes simultaneously and is blue detuned to the cavity resonance, which provides mechanical gain. The ratio between the light coupled into the cavity and the total input light is ~ 8%. The cavity transmission of the driving beam is detected by a photodetector (D2) and the signal is sent to a spectrum analyzer. The locking beam is used to stabilize the cavity resonance, which is monitored by the photodetector D1. Two weak probe beams bypass the cavity and are used to measure the motions of each membrane individually (see supplementary materials for more details). We focus on the vibrational (3,3) modes of two membranes, which are nearly degenerate with eigenfrequencies $\omega_{1,2} \sim 2\pi \times 1.2$ MHz. Piezos are used to precisely control the eigenfrequencies of membranes, hence the frequency difference of membranes $\Delta\omega = \omega_1-\omega_2$ can be either zero (i.e. completely degenerate) or as large as a few kilohertz [29,34]. In the remainder of this Letter, we denote that membrane 1 is the one with larger frequency and membrane 2 has smaller frequency.

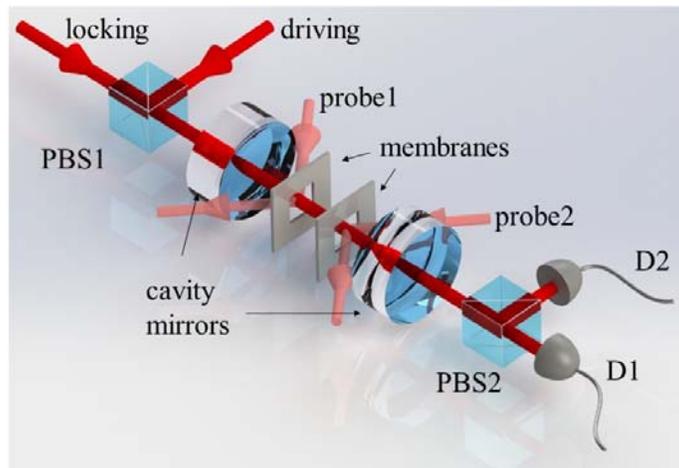

**Figure 1: Schematic diagram of the experimental setup.** Two membranes are placed inside an optical cavity. A weak locking beam is used to stabilize the cavity resonance and a strong driving beam provides mechanical gain. Two weak probe beams are used for measuring the motions of two membranes separately. PBS is the polarization beam splitter. D1 and D2 are the photodetectors for cavity locking and cavity transmission measurement, respectively.

The coupling between the membranes in such a two-membrane-in-the-middle system is through interacting with a common intracavity field, rather than direct mechanical contact. The interaction Hamiltonian of this system can be written as [25,26]

$$H_{int} = -\hbar \hat{a}^\dagger \hat{a}(G_1 \hat{q}_1 + G_2 \hat{q}_2), \qquad (1)$$

where $\hat{a}^\dagger$ ($\hat{a}$) is the creation (annihilation) operator of cavity field, $G_i$ (i=1,2) is the optomechanical coupling strength, and $\hat{q}_i$ (i=1,2) is the membrane position operator.

The equations of motion in the classical and mean field limit can thus be given by

$$\dot{\alpha} = [-\kappa/2 + i(\Delta + G_1 q_1 + G_2 q_2)]\alpha + E_{in}, \qquad (2)$$

$$m_1 \ddot{q}_1 = -m_1 \omega_1^2 q_1 - m_1 \gamma_1 \dot{q}_1 + \hbar G_1 |\alpha|^2, \qquad (3)$$

$$m_2 \ddot{q}_2 = -m_2 \omega_2^2 q_2 - m_2 \gamma_2 \dot{q}_2 + \hbar G_2 |\alpha|^2. \qquad (4)$$

Here $\alpha = \langle \hat{a} \rangle$ is the complex light amplitude and $q_{1,2} = \langle \hat{q}_{1,2} \rangle$ are the amplitudes of membrane oscillators. $m_{1,2}$ are the effective masses of membranes. $\gamma_1/2\pi$=0.53 Hz and $\gamma_2/2\pi$=0.61 Hz are the mechanical damping rates. $\Delta/2\pi$=2 MHz is the detuning of the driving laser from the cavity resonance. $\kappa/2\pi$=2 MHz is cavity decay rate. $E_{in}$ denotes the driving field amplitude.

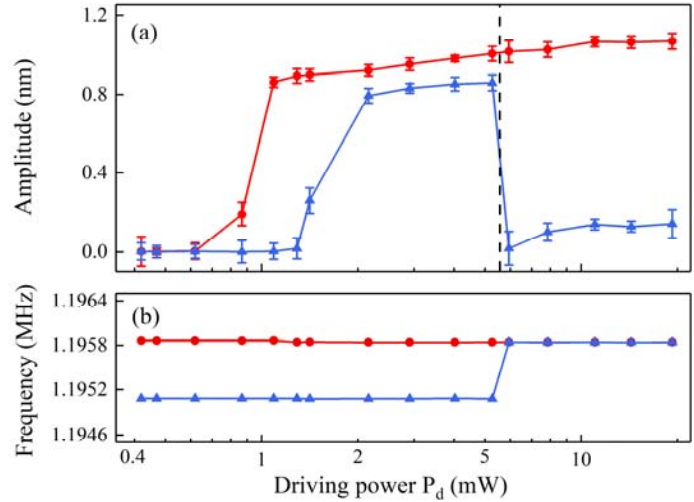

**Figure 2: Phonon lasing and synchronization transitions.** Amplitudes (a) and frequencies (b) of membrane oscillators as a function of driving power. The red circles and blue triangles are the experimental data corresponding to membrane 1 and 2, respectively. The lines are guides to the eye.

The phonon lasing and synchronization transitions are illuminated in Fig. 2. At a relatively low driving laser power, e.g. $P_d$=0.5 mW, the membranes oscillate randomly due to thermal fluctuations. When the driving power is increased, dynamical backaction amplifies the mechanical motions. The phonon lasing threshold emerges when the gain can compensate the dissipations, and the membranes oscillate coherently at their own eigenfrequencies, e.g. $P_d$=3 mW. When the driving power keeps increasing and the second threshold appears (dashed line in Fig. 2a), two membranes oscillate at the same frequency, i.e. synchronization, as shown in Fig. 2b, which can be effectively described by the Kuramoto model [35]. The oscillation amplitude of membrane 2 decreases dramatically (Fig. 2a). This is because that the natural oscillation at $\omega_2$ is suppressed and membrane 2 starts to oscillate at $\omega_1$ due to synchronization. Since the new oscillation frequency ($\omega_1$) is not close to its intrinsic mechanical resonance ($\omega_2$), the oscillation amplitude is much smaller comparing to the one at its natural frequency.

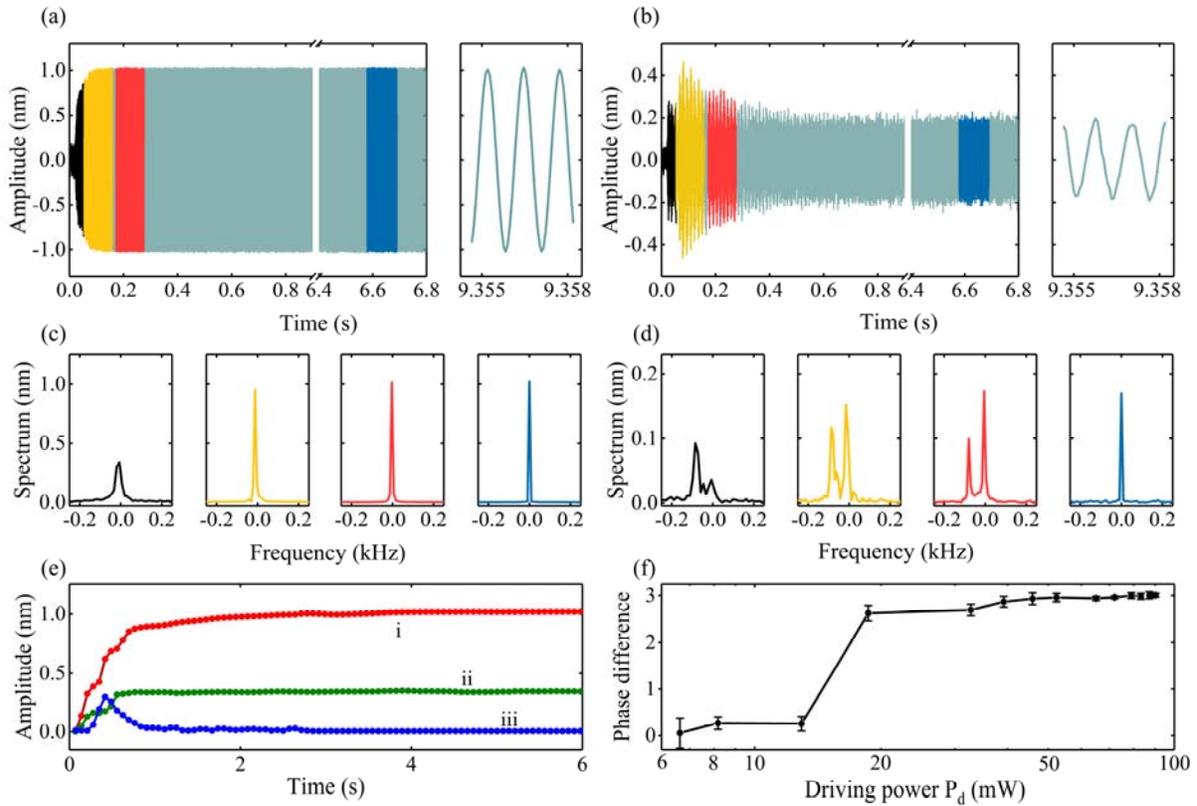

**Figure 3: Transient dynamics of self-organized synchronization.** (a) and (b) are the real-time evolution of two membranes' motions during the transient process. (c) and (d) are the Fourier transform spectra of the traces shown in (a) and (b) respectively at different time (specified by different colors). The frequency is zeroed at $\omega_1$, which is 1.19728 MHz. (e) Time evolution of amplitudes of membrane oscillators. The red curve i presents the amplitude of membrane 1 at $\omega_1$. The green curve ii and blue curve iii are the amplitudes of membrane 2 at $\omega_1$ and $\omega_2$, respectively. The curves ii and iii are multiplied by a factor of 2 for clarity. The driving power is 20 mW in (a-e). (f) In-phase to anti-phase transition. The phase difference between two membrane oscillations is plotted at different driving powers.

The dynamics of the synchronization transient process is investigated in Fig. 3. Figures 3a and 3b present the time evolution of two membranes' motions, respectively. The driving beam is abruptly switched on at t=0. When the time is long enough, two membranes are synchronized and the oscillation amplitudes become stable. Typically, the transient time gets shorter as the driving power becomes larger. As one can see in the right panels of Figs. 3a and 3b, the phase difference between two membrane oscillators is a constant, which is close to $\pi$. This is the first direct observation of phase locking between two self-organized synchronous optomechanical oscillators.

Figures 3c and 3d depict the Fourier transform spectra of membranes motions at different time. During the transient process, membrane 1 oscillates at its own eigenfrequency $\omega_1$ and the oscillation amplitude increases (Fig. 3c and red curve i in Fig. 3e). While membrane 2 oscillates at two frequencies at the beginning, i.e. the eigenfrequencies of itself ($\omega_2$) and membrane 1 ($\omega_1$, the synchronous mode). As time evolves, the oscillation amplitude at $\omega_2$ decreases (blue curve iii in Fig. 3e) and the amplitude of synchronous mode increases (green curve ii in Fig. 3e). Eventually, the oscillation at $\omega_2$ is suppressed and membrane 2 is fully synchronized to $\omega_1$ (Fig. 3d). Figure 3e clearly illustrated this transient dynamics. Such observations indicate that the system enters into the synchronization regime via torus birth bifurcation line instead of saddle-node bifurcation line [36]. It is worth noticing that the process happened on membrane 2 is very similar to the mode competition phenomenon in a multimode phonon laser [37]. By choosing different system parameters, the transient dynamics can be much more complicated, which requires further investigation. Figure 3f shows the phase difference as a function of driving power. As one can see in Fig. 3f, the system can be synchronized not only anti-phase ($\pi$ phase difference), but also in-phase (0 phase difference), which indicates two separate regimes in the phase diagram, and the transition between two regimes can be simply achieved by tuning the driving power.

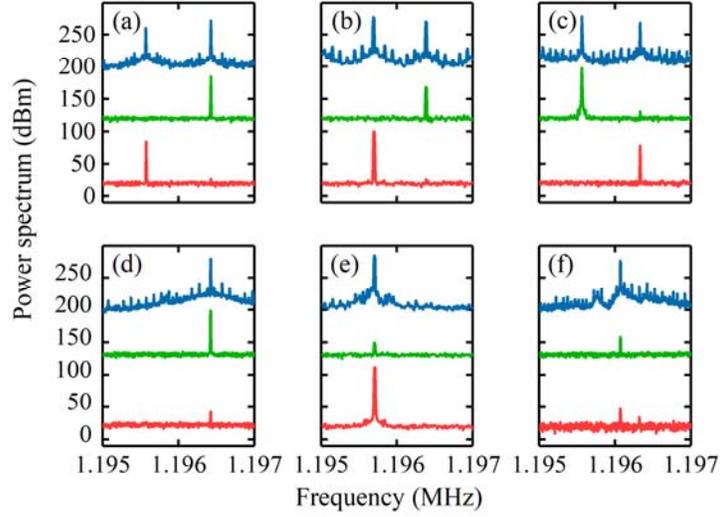

**Figure 4: Frequency entrainment of two nonidentical membrane oscillators**. (a-c) Membranes are not synchronized when the driving power is below the synchronization threshold with regard to the three different situations. (d-f) Membranes are synchronized at the eigenfrequencies of membrane 1 (d) and membrane 2 (e), and the case when the synchronous frequency is in between two eigenfrequencies (f). The blue, green, and red curves represent the power spectra of the cavity output field and two probe beams, respectively. The curves are shifted vertically for clarity. The green curve in f is multiplied by a factor of 2. $G_1/2\pi$=1.92 kHz/pm and $G_2/2\pi$=2.29 kHz/pm for (a) and (d). $G_1/2\pi$=-0.84 kHz/pm and $G_2/2\pi$=1.57 kHz/pm for (b) and (e). $G_1/2\pi$=1.23 kHz/pm and $G_2/2\pi$=2.40 kHz/pm for (c) and (f).

Such multi-membrane optomechanical systems could study a variety of synchronizations. Figure 4 shows the frequency entrainment for two nonidentical membrane oscillators at three different situations. By changing the system parameters, e.g. $G_1$, $G_2$, and $\Delta\omega$, two membrane oscillators can be synchronized at different frequencies. Figures 4d, 4e, and 4f show the cases when two membranes are synchronized at $\omega_1$, $\omega_2$, and in the between, respectively. Figures 4a-4c are the corresponding unsynchronized situations when the driving power is below the synchronization threshold. The blue, green, and red curves in Fig. 4 describe the power spectra of cavity output field and individual membranes' oscillations (measured by probe 1 and 2), respectively. The reason why the system reaches to different synchronous frequencies at various conditions in Fig. 4 can be qualitatively understood as that the frequency of the weaker oscillator is more likely to be pulled to the stronger one [2].

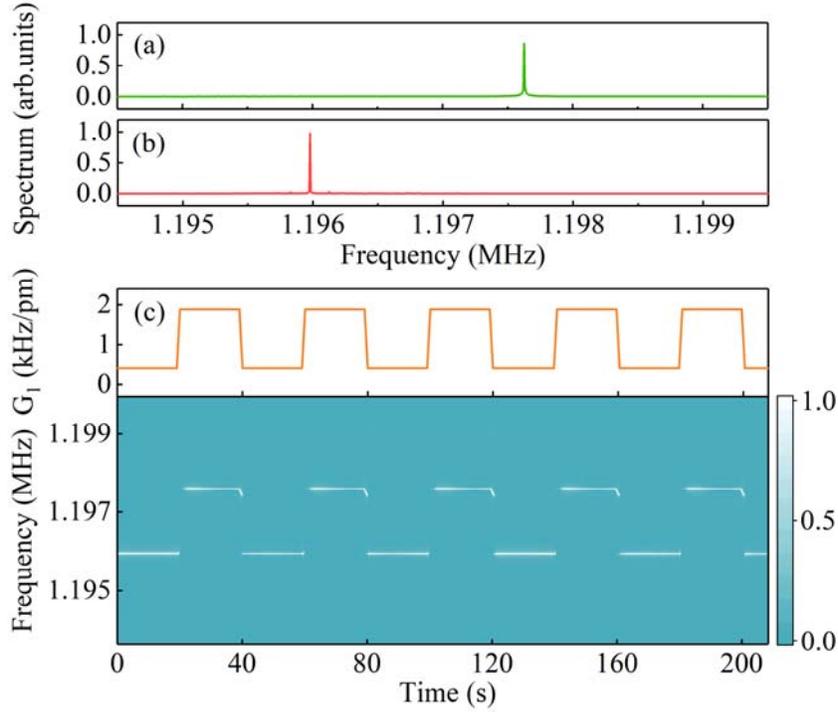

**Figure 5: Phononic digital memory.** (a) and (b) are the power spectra of cavity transmission showing oscillations at $\omega_1$=1.1976 and $\omega_2$=1.1959 MHz when $G_1$ is 1.9 and 0.4 kHz/pm, respectively. (c) The power spectrum (the lower panel) shows that the system can switch between $\omega_1$ and $\omega_2$ by dynamically modulating $G_1$ (the upper panel) with $G_2$=1.6 kHz/pm.

The controllability of the synchronization at different frequencies (as shown in Fig. 4) is used to realize a phononic digital memory, i.e. no phase coherence is associated. A controllable nanomechanical memory is essential for a possible nanomechanical computer, and has been realized in various systems [38-41], mostly based on mechanical bistability due to nonlinearities. Here synchronization is utilized as the operational mechanism. By properly choosing the parameters, the output of the system shows different power spectra. As one can see in Figs. 5(a) and 5(b), the power spectra of cavity transmission show oscillations at 1.1976 and 1.1959 MHz, which correspond to the states "1" and "0" for a bit, respectively. By periodically modulating the optomechanical coupling strength of one membrane, the oscillation frequency of the system periodically switches, as shown in Fig. 5(c). The timescale is in the order of seconds due to the long coherence time of the membranes. The transient time can be shorten by simply increasing the driving power, or maybe some more sophisticated technique, such as shortcuts [42].

In conclusion, we have demonstrated the self-organized synchronization of two membranes in an optomechanical Fabry-Perot cavity. In contrast to the previous experimental studies which are based on stable long-term measurements [30-33], our system enables to probe each membrane in real-time and individually. Consequently, the phase locking effect is directly observed, and the real-time dynamics of transient process is comprehensively studied, which might provide a new strategy to study the transient dynamics in phase transition of self-organized pattern formation and superradiance in out-of-equilibrium and complex systems [43]. Moreover, the system provides great freedom to control the optomechanical coupling. The long range interaction and dynamics can be precisely tailored by properly tuning the system parameters. As the first step to multistate switching [44] and logic gates [45,46], the phononic memory is realized by dynamically controlling the coupling strength. Although the effect has been studied is classical, the setup provides a flexible platform for studying quantum collective phenomena and classical-to-quantum transition, which provides a perspective that synchronization can be a novel tool for quantum manipulation and quantum information processing.

**Acknowledgement**

This research was supported by the National Key Research and Development Program of China (No. 2017YFA0304201), National Natural Science Foundation of China (Nos. 11925401, 11734008, 11974115, and 11704126), Program of Shanghai Subject Chief Scientist (No. 17XD1401500), the Shanghai Committee of Science and Technology (No. 17JC1400500), Natural Science Foundation of Shanghai (No. 17ZR1443100), the Shanghai Sailing Program (No. 17YF1403900), the Program for Professor of Special Appointment (Eastern Scholar) at Shanghai Institutions of Higher Learning.

**Reference**
[1] A. E. Siegman, Lasers (University Science Books, 1986).
[2] A. Pikovsky, M. Rosenblum, and J. Kurths, Synchronization: a universal concept in nonlinear sciences (Cambridge university press, 2003).

[3] J.-D. Deschenes, L. C. Sinclair, F. R. Giorgetta, W. C. Swann, E. Baumann, H. Bergeron, M. Cermak, I. Coddington, and N. R. Newbury, Synchronization of distant optical clocks at the femtosecond level, Phys. Rev. X **6**, 021016 (2016).

[4] L. Goldberg, H. F. Taylor, and J. F. Weller, Injection locking of coupled-stripe diode laser arrays, Appl. Phys. Lett. **46**, 236 (1985).

[5] L. Bartelt-Berger, U. Brauch, A. Giesen, H. Huegel, and H. Opower, Power-scalable system of phase-locked single-mode diode lasers, Appl. Opt. **38**, 5752 (1999).

[6] R. Roy and K. S. Thornburg, Experimental synchronization of chaotic lasers, Phys. Rev. Lett. **72**, 2009 (1994).

[7] G. D. Vanwiggeren and R. Roy, Communication with chaotic lasers, Science **279**, 1198 (1998).

[8] J. M.Weiner, K. C. Cox, J. G. Bohnet, and J. K. Thompson, Phase synchronization inside a superradiant laser, Phys. Rev. A **95**, 033808 (2017).

[9] I. S. Grudinin, H. Lee, O. Painter, and K. J. Vahala, Phonon laser action in a tunable two-level system, Phys. Rev. Lett. **104**, 083901 (2010).

[10] R. M. Pettit, W. Ge, P. Kumar, D. R. Luntz-Martin, J. T. Schultz, L. P. Neukirch, M. Bhattacharya, and A. Nick Vamivakas, An optical tweezer phonon laser, Nat. Photon. **13**, 402 (2019).

[11] M. Aspelmeyer, T. J. Kippenberg, and F. Marquardt, Cavity optomechanics, Rev. Mod. Phys. **86**, 1391 (2014).

[12] A. Mari, A. Farace, N. Didier, V. Giovannetti, and R. Fazio, Measures of quantum synchronization in continuous variable systems, Phys. Rev. Lett. **111**, 103605 (2013).

[13] T. E. Lee and H. R. Sadeghpour, Quantum synchronization of quantum van der Pol oscillators with trapped ions, Phys. Rev. Lett. **111**, 234101 (2013).

[14] N. Lörch, E. Amitai, A. Nunnenkamp, and C. Bruder, Genuine quantum signatures in synchronization of anharmonic self-oscillators, Phys. Rev. Lett. **117**, 073601 (2016).

[15] D. Witthaut, S. Wimberger, R. Burioni, and M. Timme, Classical synchronization indicates persistent entanglement in isolated quantum systems, Nat. Commun. **8**, 14829 (2017).

[16] L. Chang, X. Jiang, S. Hua, C. Yang, J. Wen, L. Jiang, G. Li, G. Wang, and M. Xiao, Parity–time symmetry and variable optical isolation in active–passive-coupled microresonators, Nat. Photon. **8**, 524 (2014).

[17] G. Heinrich, M. Ludwig, J. Qian, B. Kubala, and F. Marquardt, Collective dynamics in optomechanical arrays, Phys. Rev. Lett. **107**, 043603 (2011).

[18] K. E. Grutter, M. I. Davanço, and K. Srinivasan, Slot-mode optomechanical crystals: a versatile platform for multimode optomechanics, Optica **2**, 994 (2015).

[19] M. Colombano, G. Arregui, N. Capuj, A. Pitanti, J. Maire, A. Griol, B. Garrido, A. Martinez, C. Sotomayor-Torres, and D. Navarro-Urrios, Synchronization of Optomechanical Nanobeams by Mechanical Interaction, Phys. Rev. Lett. **123**, 017402 (2019).

[20] F. Massel, S. U. Cho, J.-M. Pirkkalainen, P. J. Hakonen, T. T. Heikkilä, and M. A. Sillanpää, Multimode circuit optomechanics near the quantum limit, Nat. Commun. **3**, 987 (2012).

[21] C. A. Holmes, C. P. Meaney, and G. J. Milburn, Synchronization of many nanomechanical resonators coupled via a common cavity field, Phys. Rev. E **85**, 066203 (2012).

[22] J. Thompson, B. Zwickl, A. Jayich, F. Marquardt, S. Girvin, and J. Harris, Strong dispersive coupling of a high-finesse cavity to a micromechanical membrane, Nature **452**, 72 (2008).

[23] M. Bhattacharya and P. Meystre, Multiple membrane cavity optomechanics, Phys. Rev. A **78**, 041801 (2008).

[24] M. J. Hartmann and M. B. Plenio, Steady state entanglement in the mechanical vibrations of two dielectric membranes, Phys. Rev. Lett. **101**, 200503 (2008).

[25] A. Xuereb, C. Genes, and A. Dantan, Strong coupling and long-range collective interactions in optomechanical arrays, Phys. Rev. Lett. **109**, 223601 (2012).

[26] F. Bemani, A. Motazedifard, R. Roknizadeh, M. H. Naderi, and D. Vitali, Synchronization dynamics of two nanomechanical membranes within a Fabry-Perot cavity, Phys. Rev. A **96**, 023805 (2017).

[27] P. Piergentili, L. Catalini, M. Bawaj, S. Zippilli, N. Malossi, R. Natali, D. Vitali, and G. D. Giuseppe, Two-membrane cavity optomechanics, New J. Phys. **20**, 083024 (2018).

[28] C. Gartner, J. P. Moura, W. Haaxman, R. A. Norte, and S. Groblacher, Integrated optomechanical arrays of two high reflectivity SiN membranes, Nano Lett. **18**, 7171 (2018).

[29] X. Wei, J. Sheng, C. Yang, Y. Wu, and H. Wu, Controllable two-membrane-in-the-middle cavity optomechanical system, Phys. Rev. A **99**, 023851 (2019).

[30] M. Zhang, G. S. Wiederhecker, S. Manipatruni, A. Barnard, P. McEuen, and M. Lipson, Synchronization of micromechanical oscillators using light, Phys. Rev. Lett. **109**, 233906 (2012).

[31] M. Zhang, S. Shah, J. Cardenas, and M. Lipson, Synchronization and phase noise reduction in micromechanical oscillator arrays coupled through light, Phys. Rev. Lett. **115**, 163902 (2015).

[32] M. Bagheri, M. Poot, L. Fan, F. Marquardt, and H. X. Tang, Photonic cavity synchronization of nanomechanical oscillators, Phys. Rev. Lett. **111**, 213902 (2013).

[33] E. Gil-Santos, M. Labousse, C. Baker, A. Goetschy, W. Hease, C. Gomez, A. Lematre, G. Leo, C. Ciuti, and I. Favero, Light-mediated cascaded locking of multiple nano-optomechanical oscillators, Phys. Rev. Lett. **118**, 063605 (2017).

[34] S. Wu, J. Sheng, X. Zhang, Y. Wu, and H. Wu, Parametric excitation of a SiN membrane via piezoelectricity, AIP Advances **8**, 015209 (2018).

[35] J. A. Acebron, L. L. Bonilla, C. J. Perez Vicente, F. Ritort, and R. Spigler, The Kuramoto model: A simple paradigm for synchronization phenomena, Rev. Mod. Phys. **77**, 137 (2005).

[36] A. Balanov, N. Janson, D. Postnov, and O. Sosnovtseva, Synchronization: From simple to complex, Springer, 2008.

[37] U. Kemiktarak, M. Durand, M. Metcalfe, and J. Lawall, Mode competition and anomalous cooling in a multimode phonon laser, Phys. Rev. Lett. **113**, 030802 (2014).

[38] R. L. Badzey, G. Zolfagharkhani, A. Gaidarzhy, and P. Mohanty, A controllable nanomechanical memory element, Appl. Phys. Lett. **85**, 3587 (2004).

[39] I. Mahboob and H. Yamaguchi, Bit storage and bit flip operations in an electromechanical oscillator, Nat. Nanotechnol. **3**, 275 (2008).

[40] W. J. Venstra, H. J. R. Westra, and H. S. J. van der Zant, Mechanical stiffening, bistability, and bit operations in a microcantilever, Appl. Phys. Lett. **97**, 193107 (2010).

[41] M. Bagheri, M. Poot, M. Li, W. P. H. Pernice, and H. X. Tang, Dynamic manipulation of nanomechanical resonators in the high-amplitude regime and non-volatile mechanical memory operation, Nat. Nanotechnol. **6**, 726 (2011).

[42] S. Deng, P. Diao, Q. Yu, A. del Campo, and H. Wu, Shortcuts to adiabaticity in the strongly coupled regime: Nonadiabatic control of a unitary Fermi gas, Phys. Rev. A **97**, 013628 (2018).

[43] A. Arenas, A. Díaz-Guilera, J. Kurths, Y. Moreno, C. Zhou, Synchronization in complex networks. Phys. Rep. **469**, 93 (2008).

[44] J. Sheng, U. Khadka, and M. Xiao, Realization of all-optical multistate switching in an atomic coherent medium, Phys. Rev. Lett. **109**, 223906 (2012).

[45] I. Mahboob, E. Flurin, K. Nishiguchi, A. Fujiwara, and H. Yamaguchi, Interconnect-free parallel logic circuits in a single mechanical resonator, Nat. Commun. **2**, 198 (2011).

[46] F. Li, P. Anzel, J. Yang, P. G. Kevrekidis, and C. Daraio, Granular acoustic switches and logic elements, Nat. Commun. **5**, 5311 (2014).